\documentclass[aps,prb,twocolumn,floatfix,showpacs,citeautoscript,longbibliography,hyperlinks,superscriptaddress]{revtex4-2}

\usepackage{amsmath}
\usepackage{epstopdf}
\usepackage{amsmath}
\usepackage{amssymb}
\usepackage{amsfonts}
\usepackage[colorlinks=true,citecolor=blue]{hyperref}
\usepackage{graphicx,graphics}
\usepackage{graphicx}
\usepackage{bm}
\usepackage{hyperref}
\usepackage{braket,bbold,color}
\usepackage{tcolorbox}
\usepackage{ifthen}
\usepackage{dsfont}
\usepackage{MnSymbol}
\usepackage{graphicx}
\usepackage{placeins}
\usepackage{xcolor}

\newcommand{\revision}[1]{{{#1}}}

\newcommand{\coef}{\lambda}
\newcommand{\Ham}{\mathcal{H}}

\begin{document}

\title{Lee-Yang formalism for phase transitions of interacting fermions using tensor networks} 

\author{Pascal M. Vecsei}
\affiliation{Department of Applied Physics, Aalto University, 00076 Aalto, Finland}

\author{Jose L. Lado}
\affiliation{Department of Applied Physics, Aalto University, 00076 Aalto, Finland}

\author{Christian Flindt}
\affiliation{Department of Applied Physics, Aalto University, 00076 Aalto, Finland}
\affiliation{RIKEN Center for Quantum Computing, Wakoshi, Saitama 351-0198, Japan}

\begin{abstract}
Predicting the phase diagram of interacting quantum many-body systems is a challenging problem in condensed matter physics. Strong interactions and correlation effects may lead to exotic states of matter, such as quantum spin liquids and unconventional superconductors, that often compete with other symmetry broken states including ordered magnets and charge density waves. Here, we put forward a formalism for determining the phase diagram of fermionic systems that combines recent progress in the field of Lee-Yang theory of phase transitions with many-body tensor-network methods. Using this strategy, we map out the phase diagram of a fermionic chain, where charge density waves form owing to strong repulsion. Specifically, from the high cumulants of the order parameter, we extract the dominant zeros of the moment-generating function in chains of finite size. By extrapolating their positions to the thermodynamic limit, we determine the boundaries between competing phases. Our formalism provides a strategy for determining critical points in fermionic systems, and it is based on fluctuations of the order parameter, which are measurable quantities.
\end{abstract}

\maketitle

\section{Introduction}
\label{sec:introduction}

Determining the phase diagram of interacting many-body systems is crucial for understanding quantum matter. A variety of exotic states, including correlated superconductors~\cite{anderson1997thetheory,Qin2022,Arovas2022}, quantum spin liquids~\cite{zhou2017quantum,imada2021diractype, japaridze2007phases}, and quantum magnets~\cite{iqbal2016spin}, exhibit strong correlations that typically require exact many-body solutions~\cite{kitaev2006anyons, ghosh2023maple}. However, those solutions are often restricted to small systems, and generic quantum many-body systems still pose a formidable challenge, even for advanced numerical approaches. Tensor-network methods, for instance, deliver accurate results in one dimension, but become computationally demanding in higher dimensions~\cite{Schollwck2011thedensity}. In addition, quantum Monte Carlo methods suffer from the notorious sign problem, which prevents them from being applied to fermionic systems and frustrated magnets~\cite{sorella2017quantum,foulkes2001quantum}. Given these challenges, the development of alternative strategies for determining the phase diagrams of quantum many-body systems has become essential for advancing the field of correlated quantum matter~\cite{PhysRevX.5.041041}.

\begin{figure}
 \centering
 \includegraphics[width = 0.98\columnwidth]{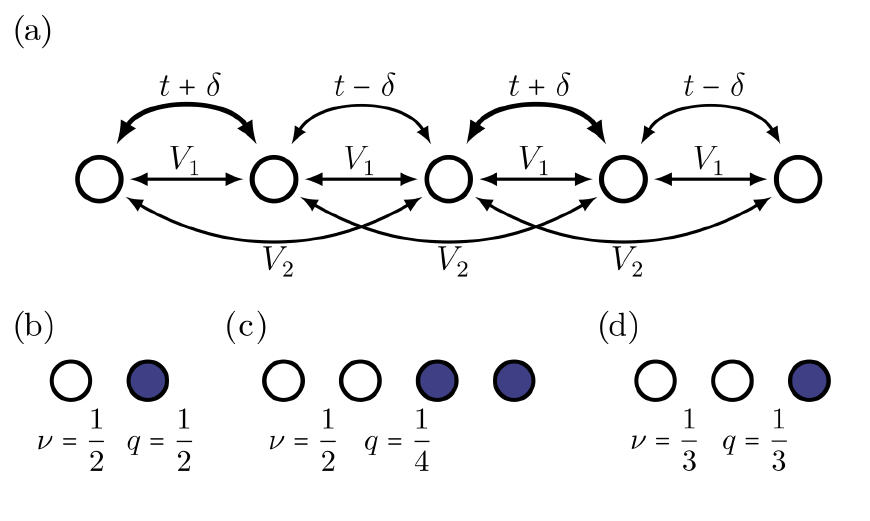}
 \caption{Fermionic chain and charge density waves. (a)~Fermionic chain with hopping amplitude $t$, dimerization~$\delta$, and nearest-neighbor and next-nearest-neighbor interactions $V_1$ and $V_2$, respectively. The system exhibits phase transitions between charge density waves (CDWs) and phases with a uniform charge distribution. (b)-(d) Illustrations of CDWs that can appear, where $\nu$ denotes the filling factor of the chain, and~$q$ is the expected wave vector of the CDW.
}
\label{fig:introPlot}
\end{figure}

Equilibrium phase transitions can be understood through 
the Lee-Yang zeros of the partition function in the complex plane of the control parameter, such as an applied magnetic field or the inverse temperature~\cite{leeYang1952StatisticalI,leeYang1952StatisticalII,fisher1965nature,blythe2003theleeyang,bena2005statistical}. As the system size increases, the zeros approach the points on the real axis, where phase transitions occur. In recent years, these ideas have been extended to a wide range of nonequilibrium situations, including phase transitions in quantum many-body systems after a quench~\cite{heyl2013dynamical,Zvyagin2016,peotta2020determination,brange2022dynamical} and dynamical phase transitions in glass formers~\cite{Hedges2009,Flindt2013}. In these nonequilibrium settings, the role of the partition function is played by the Loschmidt amplitude or by moment-generating functions, whose complex zeros signal the occurrence of phase transitions in the thermodynamic limit. In addition to these theoretical developments, Lee-Yang zeros have been determined in several recent experiments~\cite{Binek1998,Wei2012,Peng2015,Brandner2017,Francis2021,Matsumoto2022,Gao2024}.

The Lee-Yang formalism of phase transitions has been recently further developed~\cite{kist2021leeyang,vecsei2022leeyang,vecsei2023leeyang} by extending earlier studies on equilibrium phase transitions~\cite{deger2018leeYang, deger2019determination, deger2020leeYang, deger2020leeYang2} to the quantum realm. In this approach, the central quantities are the complex values for which the moment-generating function of the order parameter vanishes. The dominant zeros are extracted from the high cumulants of the order parameter, which can be evaluated numerically or even measured in experiments~\cite{Brandner2017}. By extrapolating the position of the zeros to the thermodynamic limit, the boundaries between competing phases can be identified. So far, this approach has focused on lattices of interacting spins~\cite{kist2021leeyang,vecsei2022leeyang,vecsei2023leeyang}. However, implementing it for interacting fermions~\cite{yamamoto2009dense,fodor2019trying} would be the next natural development for this Lee-Yang formalism. 

In this paper, we apply the Lee-Yang formalism to interacting fermions. Specifically, we use it to construct the phase diagram of the interacting fermionic chain in Fig.~\ref{fig:introPlot}(a) using tensor-network calculations of the high cumulants of the order parameter. In particular, we are interested in the charge density waves (CDWs) illustrated in Figs.~\ref{fig:introPlot}(b)-\ref{fig:introPlot}(d), which may form if the interactions are strong. \revision{To implement our Lee-Yang formalism, we extend it in several ways:} For each type of CDW, we \revision{need to} identify a suitable order parameter, which \revision{may be non-Hermitian}. We also exploit symmetries of the system, which directly translate into symmetries of the moment-generating function and the position of its zeros. \revision{Finally, as we will see, two, three, or four equidistant zeros of the moment generating function may approach the origin of the complex plane, which affects the high cumulants in different ways. For that reason, the zeros must be extracted from the high cumulants using different methods, depending on the number of dominant zeros.} Our results provide a starting point for applying our Lee-Yang formalism to other symmetry-breaking quantum phase transitions in interacting fermionic systems.

The rest of our paper is organized as follows. In Sec.~\ref{sec:models}, we describe the model of an interacting fermionic chain, which will be our main focus. In particular, we will investigate the emergence of CDWs owing to strong interactions. In Sec.~\ref{sec:leeYangCorrFerm}, we introduce the general principles of the Lee-Yang formalism. This approach allows us to predict phase transitions from the high cumulants of the order parameter, which we evaluate using tensor-network methods. In Sec.~\ref{sec:ssh}, we use the formalism to locate phase transitions in the fermionic chain, exploiting the symmetries of the system and the position of the zeros. In Sec.~\ref{eq:PDs}, we present the phase diagram of the system. Finally, in Sec.~\ref{sec:conclusion}, we state our conclusions. Some technical details are presented in the appendices.

\section{Interacting Fermionic chain}
\label{sec:models}

As illustrated in Fig.~\ref{fig:introPlot}(a), we consider a chain of fermions with dimerized hopping as well as nearest- and next-nearest neighbor interactions. It can be understood as a combination of the Su-Shrieffer-Heger (SSH) model with nearest-neighbor repulsion~\cite{yahyavi2018variational, melo2023topological, xianglong2020topological} and a fermionic chain with uniform hopping and nearest- and next-nearest-neighbor repulsion~\cite{zhuravlev2000breakdown, duan2011bondorder, mishra2011phasediagram, szyniszewski2018fermionic, gotta2021pairing}. The SSH model was originally developed to describe polyacetylene~\cite{su1980soliton}, and it has turned out to be a paradigmatic example of a topological insulator. It can also be experimentally realized in artificial systems, such as chains of adatoms~\cite{Drost2017,pham2022topological}. 

The Hamiltonian \revision{of} the system reads
\begin{equation}
 \hat\Ham =  \sum_{j=1}^L\left[ t_j (\hat c_j^\dag \hat c_{j+1} + \hat c_{j+1}^\dag \hat c_j) \\ 
 + V_1  \hat n_j \hat n_{j+1} + 
 V_2  \hat n_j \hat n_{j+2}\right],
\end{equation}
where $\hat c_j$ and $\hat c_{j}^\dag$ are fermionic annihilation and creation operators for site $j=1\ldots,L$ with the number operator~$\hat n_j = \hat c_j^\dag \hat c_{j}$. The hopping amplitude between neighboring sites is denoted by
$t_j=t  - (-1)^j \delta$, and it alternates between $t-\delta$ and $t+\delta$ along the chain. We have also included nearest-neighbor and  next-nearest-neighbor repulsion of strength $V_1$ and 
$V_2$. We impose periodic boundary conditions by setting $\hat c_{L+1}=\hat c_1$ and $\hat c_{L+2}=\hat c_2$, and we consider only even chain lengths to ensure that the hopping amplitude can alternate throughout the closed chain. 

The system is known to exhibit a variety of CDWs, which depend on the dimerization of the hopping amplitude, $\delta$, the interaction strengths,  $V_{1,2}$, and the filling factor of the chain, which we denote by $\nu$~\cite{yahyavi2018variational, melo2023topological, xianglong2020topological, zhuravlev2000breakdown, duan2011bondorder, mishra2011phasediagram, szyniszewski2018fermionic, gotta2021pairing}.  As shown in Fig.~\ref{fig:introPlot}(b)-\ref{fig:introPlot}(d), each of these phases can be characterized by a charge modulation with period $p$ and corresponding wave vector $q = 1/p$. For example, at half filling, $\nu=1/2$, and no next-nearest-neighbor repulsion, $V_2 = 0$, the system exhibits a CDW with $q = 1/2$~\cite{yahyavi2018variational, melo2023topological, xianglong2020topological}. Similarly, there can be CDWs with $q=1/4$  for $\nu=1/2$, and $q=1/3$ for $\nu=1/3$~\cite{gotta2021pairing}. 

\section{Lee-Yang formalism}
\label{sec:leeYangCorrFerm}

\begin{figure*}
 \centering
 \includegraphics[width = 1.0\textwidth]{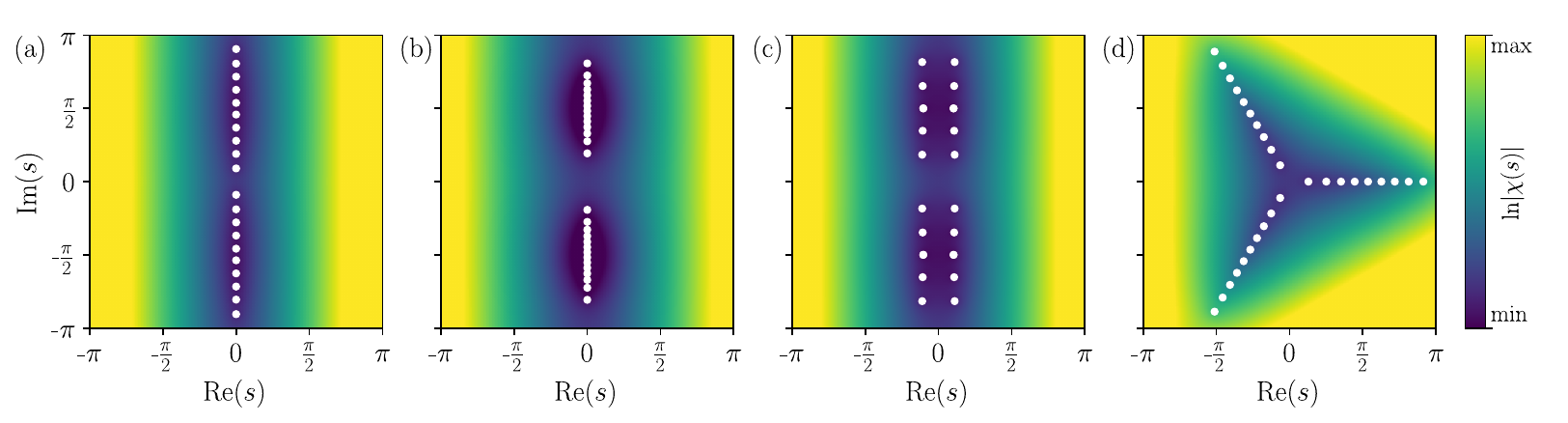}
 \caption{\label{fig:heatmapPlotSplane}Complex zeros of the moment-generating function. We show the logarithm of the moment-generating function, $\ln |\chi(s)|$, and indicate the  zeros by white dots as found by a minimum search algorithm. (a) Complex zeros for $\nu=1/2$, $\delta = 0.5$, $V_1 = 5$, $V_2 = 0$ and $L=20$, corresponding to a CDW with $q=1/2$. (b) Complex zeros for $\nu=1/2$, $\delta = 0$, $V_1 = 1$, $V_2 = 0$, and $L = 28$, corresponding to a CDW with $q=1/2$.  (c) Complex zeros for $\nu=1/2$, $\delta = 0$, $V_1 = 0$, $V_2 = 2$ and $L=20$, corresponding to a CDW with $q=1/4$. (d) Complex zeros for $\nu=1/3$, $\delta = 0$, $V_1 = 7$, $V_2 = 2$, and  $L=30$, corresponding to a CDW with $q=1/3$.}
\end{figure*}

The central idea of the Lee-Yang formalism that we employ is that the CDWs can be identified from the complex zeros of the moment-generating function that characterizes the fluctuations of the order parameter~\cite{kist2021leeyang, vecsei2022leeyang, vecsei2023leeyang}. \revision{Originally, Lee and Yang developed their formalism to understand phase transitions in classical physics~\cite{leeYang1952StatisticalI,leeYang1952StatisticalII,fisher1965nature,blythe2003theleeyang,bena2005statistical}. To do so, they considered the zeros of the partition function in the complex plane of the control parameter, which, for example, can be a magnetic field, the inverse temperature, or the fugacity. For systems of finite size, the zeros are complex. However, as the system size is increased, the zeros approach the point on the real axis, where a phase transition occurs. For quantum systems at zero temperature, one may consider the zeros of moment-generating functions, which then play the role of the partition function~\cite{kist2021leeyang, vecsei2022leeyang, vecsei2023leeyang}.} As we will see, the dominant zeros \revision{of the moment-generating function} can be extracted from the high cumulants of the order parameter, which in principle are measurable. Below, we define the order parameters of the CDWs, but our approach may also be applied to other types of phases and order parameters. For the CDWs, we take order parameters of the form
\begin{equation}
\label{eq:generalOrderParameterCDW}
 \hat O_{q} = \sum_j \coef_j \hat n_j,
\end{equation}
where the choice of coefficients $\coef_j$ depend on the corresponding CDW. For example, for $q=1/2$ as in~Fig.~\ref{fig:introPlot}(b), we take $\coef_j = (-1)^j$, and the order parameter becomes
\begin{equation}
\label{eq:OPq1/2}
 \hat O_{1/2} = \sum_j (-1)^j \hat n_j,
\end{equation}
which is nonzero for that particular CDW. Similarly for $q = 1/4$ as in~Fig.~\ref{fig:introPlot}(c), we use coefficients $\coef_j$ with the alternating series $\ldots, -1, -1, +1, +1, \ldots$, while for $q = 1/3$ as in~Fig.~\ref{fig:introPlot}(d), we take $\coef_j = \exp \left( i 2 \pi j / 3 \right)$. 

We now define the moment-generating function of the order parameter as the ground state average
\begin{equation}
 \chi(s) = \langle e^{s \hat O_{q } } \rangle =  \sum_{k=1}^M \langle \Psi_k|e^{s \hat O_{q}}|
 \Psi_k\rangle/M,
 \label{eq:mgen}
\end{equation}
where we have included the possibility that the ground state may be $M$-fold degenerate. We can then obtain the $n$th moment of the order parameter as the $n$th derivative with respect to the counting variable $s$ at $s=0$,
\begin{equation}
 \langle \hat{O}_q^n \rangle =  \partial_s^n \chi(s)\big|_{s=0}. 
 \label{eq:moments}
\end{equation}
We also define the cumulant generating function,
\begin{equation} 
\Theta(s) = \ln \chi(s),
\end{equation}
whose derivatives yield the cumulants as
\begin{equation}
 \llangle \hat{O}_q^n \rrangle =  \partial_s^n \Theta(s)\big|_{s=0}.
\end{equation}
The first two cumulants are the average, $\llangle \hat{O}_q\rrangle =  \langle \hat{O}_q\rangle$, and the variance, $\llangle \hat{O}_q^2\rrangle =  \langle \hat{O}_q^2\rangle- \langle \hat{O}_q\rangle^2$. \revision{Generally, the cumulants can be expressed in terms of the moments as
\begin{equation}
\langle\!\langle \hat{O}_q^n \rangle\!\rangle=\langle \hat{O}_q^{n} \rangle-\sum_{m=1}^{n-1} \binom{n-1}{m-1} \langle\!\langle \hat{O}_q^m \rangle\!\rangle \langle \hat{O}_q^{n-m} \rangle.
\label{eq:mom2cumu}
\end{equation}
The variance is non-negative, while the higher cumulants, such as the skewness, $\langle\!\langle \hat{O}_q^3 \rangle\!\rangle$, and the kurtosis, $\langle\!\langle \hat{O}_q^4 \rangle\!\rangle$, can be either positive or negative. In the following, we will make systematic use of the high cumulants beyond  the skewness and the kurtosis. As we will see, they can be used to pinpoint the location of a phase transition.}

We now make use of the property that the moment generating function for finite-size systems is an entire function, which can be expressed in terms of its zeros as
\begin{equation}
 \chi(s) = e^{cs} \prod_k \left( 1 - s/s_k \right).
\label{eq:productExpressionForMomGenFUn}
\end{equation}
Here, the complex zeros are denoted by $s_k$, while $c$ is a constant, which will not be important in the following.
Moreover, the number of zeros and their location in the complex plane depend on the system size. Importantly, phase transitions can be detected from the motion of the zeros in the complex plane as the system size increases~\cite{kist2021leeyang, vecsei2022leeyang, vecsei2023leeyang}. Indeed, in the thermodynamic limit, the order parameter is expected to be nonanalytic if the system exhibits a phase transition. This nonanalytic behavior emerges because the zeros of the moment-generating function approach the origin of the complex plane, where the derivatives are evaluated to obtain the moments and the  cumulants~\cite{kist2021leeyang}. Away from the phase transition, the zeros do not reach the origin in the thermodynamic limit. 

\begin{figure*}
 \centering
 \includegraphics[width = 1.0\textwidth]{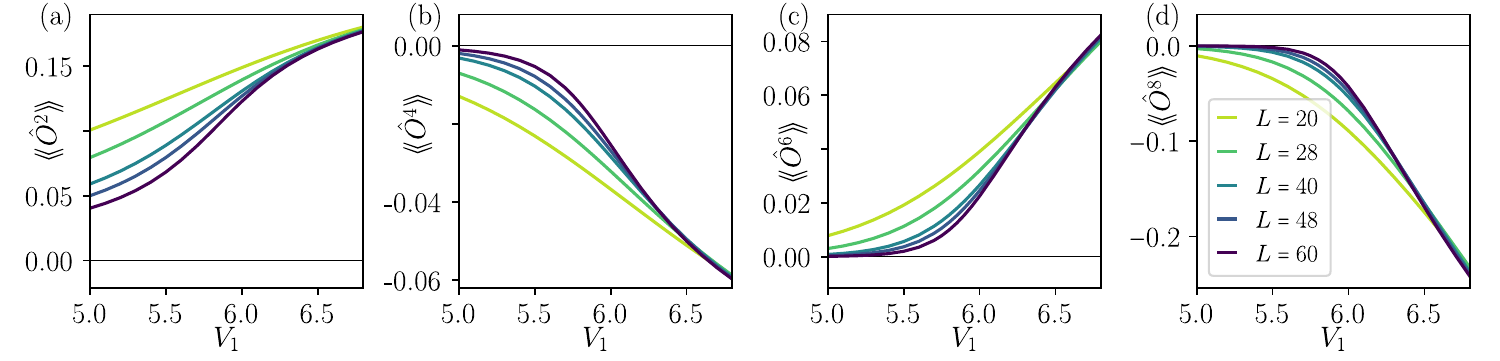}
 \caption{\label{fig:cumulants} \revision{High cumulants. (a)-(d) Second, fourth, sixth, and eight cumulant of $\hat O_{1/2}$ as functions of $V_1$ with $\delta = 0.5$ and $V_2 = 0$ for different system sizes $L$. Because of symmetries, the odd cumulants vanish according to Eq.~(\ref{eq:approxcumusinglepair}), and they are not shown here.}}
 \label{fig:cumu}
\end{figure*}

In Fig.~\ref{fig:heatmapPlotSplane}, we show examples of the complex zeros of the moment-generating function. To this end, we plot the logarithm of the moment-generating function with the zeros indicated by white dots. We consider a chain of finite size, with each panel corresponding to different parameter values. As we will now see, we can determine the zeros that are closest to the origin from the high cumulants of the order parameter for finite-size systems. We then extrapolate their position with increasing system size to find their convergence points in the thermodynamic limit and thereby predict phase transitions.  To find the zeros that are closest to the origin, we express the cumulants in terms of the zeros. Specifically, from Eq.~(\ref{eq:productExpressionForMomGenFUn}), we see that the cumulant-generating function can be written as 
\begin{equation}
\Theta(s) = cs+\sum_k \ln ( 1 - s/s_k) ,
\end{equation}
and the cumulants then become
\begin{equation}
 \llangle \hat O_q^n \rrangle =  - \sum_k (n-1)!/s_k^n,\quad n>1.
\label{eq:cumulantsAsLYZsum}
\end{equation}
From this expression, we see that the high cumulants are dominated by the zeros that are closest to the origin, since the relative contribution from other zeros is suppressed as the cumulant order is increased. Thus, for sufficiently high orders, the cumulants are determined by the few zeros that are closest to the origin. Moreover, as we will see, we can extract these zeros from the high cumulants. To evaluate the high cumulants, we use tensor-network calculations as detailed in~Appendix~\ref{app:mpsdetails}. After having extracted the zeros for different system sizes, we extrapolate their position to the thermodynamic limit. If the zeros reach the origin of the complex plane, the system is at a phase transition. This approach then allows us to map out the phase diagram of the system.

\section{Symmetries and zeros} 
\label{sec:ssh}

We first consider the system without next-nearest-neighbor interactions, $V_2=0$, and take the chain to be half-filled with a finite dimerization. With increasing nearest-neighbor interactions, we then expect a CDW to develop, and we now use the Lee-Yang formalism to investigate this transition. For this CDW with $q=1/2$, we use the order parameter in Eq.~(\ref{eq:OPq1/2}). To begin with, we identify several symmetries of the moment-generating function, which constrain the positions of the zeros. First, we note that the system is inversion symmetric with respect to the center of the chain with an even number of sites, since the Hamiltonian fulfils
\begin{equation}
\hat{\mathcal S} \hat \Ham \hat{\mathcal S} ^\dag = \hat \Ham, 
\end{equation}
where $\hat{\mathcal S}$ is the operator for inversion symmetry with $\hat{\mathcal S} \hat{\mathcal S}^\dagger= 1$.
We can now consider the moment-generating function, here at a finite temperature,
\begin{equation}
\chi_\beta(s) =  \text{tr}\{ e^{-\beta \hat \Ham}e^{s \hat O_{q}}\}/Z_\beta,
\end{equation}
where $Z_\beta=\text{tr}\{ e^{-\beta \hat \Ham}\}$ is the equilibrium partition function at the inverse temperature $\beta$. We then find
\begin{equation}
\begin{split}
\chi_\beta(s) &=  \text{tr}\{ e^{-\beta \hat{\mathcal S} \hat \Ham  \hat{\mathcal S}^\dag} e^{s \hat O_q}\} /Z_\beta\\
&=  \text{tr}\{ e^{-\beta \hat \Ham} e^{s \hat{\mathcal S}^\dag \hat O_q \hat{\mathcal S}}\}/Z_\beta= \chi_\beta(-s), 
\end{split}
\end{equation}
where, in the last step, we have used that both $\hat O_{q=1/2}$ and $\hat O_{q=1/4}$ fulfill the symmetry
\begin{equation}
\hat{\mathcal S}^\dag \hat O_q \hat{\mathcal S}=-\hat O_q.
\end{equation}
The moment-generating function is then point-symmetric with respect to the origin of the complex plane, also at zero temperature, which is relevant here.

We find an additional symmetry at half filling for chain lengths that are a multiple of four. In that case, the eigenvalues of $\hat O_{q=1/2}$ and $\hat O_{q = 1/4}$ are even integers, and
\begin{equation}
 \chi(s \pm i\pi) =  \langle e^{s \hat O_{q }}e^{\pm i\pi \hat O_{q }} \rangle
 = \chi(s),
\end{equation}
This symmetry implies that the zeros are periodic in the direction of the imaginary axis. Thus, if $s_k$ is a zero of the moment-generating function, then $s_k + i\pi$ is also a zero.  Finally, for Hermitian operators, we can write 
\begin{equation}
 \chi(s) = \langle e^{s \hat O_q} \rangle =  \sum_j  p_{j}  e^{sO_j},
\end{equation}
where the sum runs over all eigenvalues of $\hat O_q$, and $p_j$ is the probability of measuring the eigenvalue $O_j$. Since $p_j$ and $O_j$ are both real, we  find
\begin{equation}
 \chi^*(s)  = \sum_j p_j e^{s^* O_j} = \chi(s^*),
\end{equation}
such that the zeros are mirrored across the real axis. Thus, if $s_k$ is a zero, then $s^*_k$ is also a zero. By combining these symmetries, we find that the zeros must come as pairs on the imaginary axis or as four zeros on the corners of a rectangle centered at the origin. Such configurations are seen in Figs.~\ref{fig:heatmapPlotSplane}(a)-\ref{fig:heatmapPlotSplane}(c), while Fig.~\ref{fig:heatmapPlotSplane}(d) corresponds to a CDW with $q=1/3$, which we will return to. 

\begin{figure*}
	\centering
	\includegraphics[width = 1.0\textwidth]{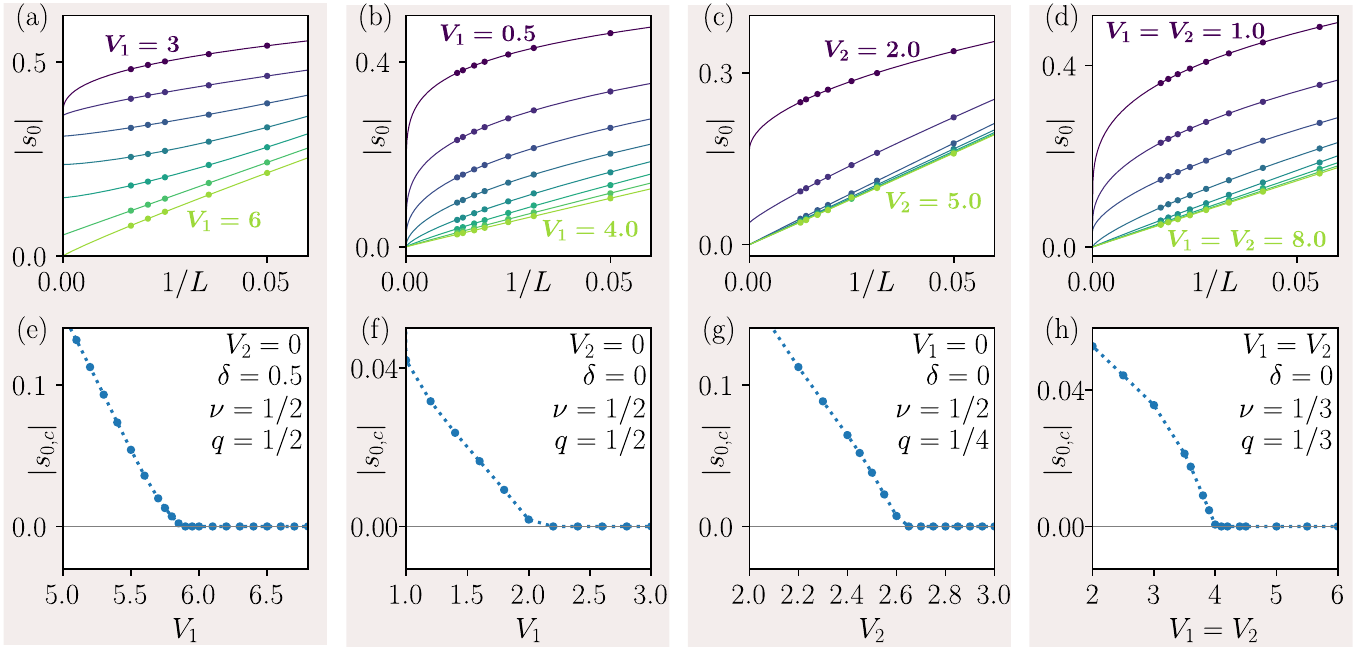}
	\caption{\label{fig:lyzScaling} Extraction of zeros. (a) Extracted zeros as a function of the system size for the CDW with $q=1/2$. Parameters are $\nu=1/2$, $\delta = 0.5$, and $V_2 = 0$. The lines show a power-law extrapolation to the thermodynamic limit.  (b) Similar results for the CDW with $q=1/2$ and the parameters $\nu=1/2$, $\delta = 0$, and $V_2 = 0$. 
		(c) Results for the  CDW with $q=1/4$ and the parameters $\nu=1/2$,  $\delta = 0$, and $V_1 =0$. (d) Results for the  CDW with $q=1/3$ and the parameters $\nu=1/3$, $\delta = 0$, and  $V_1 = V_2$. (e)-(h)~Convergence point in the thermodynamic limit corresponding to the results in panels (a)-(d). }
\end{figure*}

We first consider the situation in Fig.~\ref{fig:heatmapPlotSplane}(a), where a single pair of zeros are closest to the origin. In that case, we can approximate Eq.~(\ref{eq:cumulantsAsLYZsum}) as
\begin{equation}
 \llangle \hat O_q^n \rrangle \simeq  - (n-1)!(1+(-1)^{n})/i^{n}|s_0|^n, \quad n\gg 1
 \label{eq:approxcumusinglepair}
\end{equation}
since the contributions from zeros that are further away are suppressed exponentially with the cumulant order. Thus, the odd cumulants vanish, while the even ones read
\begin{equation}
 \llangle \hat O_q^{2n} \rrangle \simeq  - (2n-1)! 2(-1)^{n}/|s_0|^{2n}.
\end{equation}
We can then express the closest pair of zeros as
\begin{equation}
  |s_0| \simeq\left[ (2n-1)!2/ |\llangle \hat O_q^{2n} \rrangle|\right]^{1/2n}.
\label{eq:zerosextractionpair}
\end{equation}
in terms of the even cumulants. Thus, from the high cumulants, we can extract the position of the zeros and follow their motion as we increase the system size. 

\revision{In Fig.~\ref{fig:cumu}, we show calculations of the high cumulants corresponding to the situation in Fig.~\ref{fig:heatmapPlotSplane}(a). We then use these cumulants to extract the zeros using Eq.~(\ref{eq:zerosextractionpair}).} In Fig.~\ref{fig:lyzScaling}(a), we show the results of this procedure for different values of the nearest-neighbor interaction and different chain lengths. We also show a power-law extrapolation to the thermodynamic limit as indicated by a line. We can thereby determine the convergence point in the thermodynamic limit, which we show in Fig.~\ref{fig:lyzScaling}(e) as a function of the nearest-neighbor interaction strength. If the zeros converge to the origin of the complex plane, the system is in the ordered phase, where a CDW exists. On the other hand, if the zeros do not reach the origin, the system is in the disordered phase without a CDW. Thus, from Fig.~\ref{fig:lyzScaling}(e), we can determine the value of the nearest-neighbor interaction for which the transition between the two phases occurs.

Next, we consider the system with vanishing dimerization, either at half filling or at one-third filling. At one-half filling, we expect CDWs  to develop with either $q=1/2$ or $q = 1/4$~\cite{zhuravlev2000breakdown, duan2011bondorder, mishra2011phasediagram, szyniszewski2018fermionic, gotta2021pairing}, while at one-third filling, we expect a CDW phase with $q= 1/3$ to appear as the interactions are increased~\cite{gotta2021pairing}. We first focus on the CDW with $q = \frac{1}{2}$ for one-half filling. An example of the zeros in the complex plane is shown in Fig.~\ref{fig:heatmapPlotSplane}(b). In  Fig.~\ref{fig:lyzScaling}(b), we show the zeros that we extract from the high cumulants for different values of the nearest-neighbor interaction and different chain lengths.  We also indicate the extrapolation
to the thermodynamic limit by a line. In Fig.~\ref{fig:lyzScaling}(f), we show the convergence point as a function of the  nearest-neighbor interaction, and we can again identify the value for which a phase transition occurs.

\begin{figure*}
 \centering
 \includegraphics[width = 1.0\textwidth]{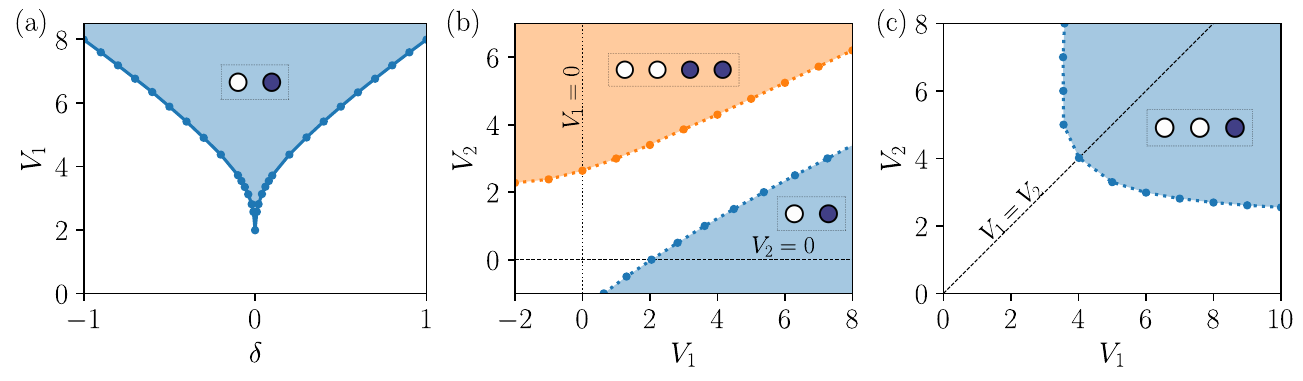}
 \caption{\label{fig:phaseDiagrams} Phase diagrams. (a) Phase diagram at half filling and $V_2=0$.  A CDW with $q=1/2$ develops in the blue region. (b)~Phase diagram at half filling and $\delta=0$. A CDW with $q=1/2$ develops in the blue region, while  a CDW with $q=1/4$ forms in the orange region. (c) Phase diagram at one-third filling
 and $\delta=0$. A CDW with $q=1/3$ appears in the blue region.}
\end{figure*}

We now increase the next-nearest-neighbor interactions, which leads to a CDW with $q=1/4$ to develop. In this case, we use the order parameter $\hat O_{q}= \sum_{j=1}^L \coef_j \hat n_j$, where the coefficients $\coef_j$ are given by  the alternating  series $\ldots, -1, -1, +1, +1, \ldots$. An example of the corresponding zeros is shown in Fig.~\ref{fig:heatmapPlotSplane}(c), where we see that the zeros are no longer located on the imaginary axis, but rather arranged on the corners of a rectangle.  For sufficiently high cumulant orders, only the four zeros that are closest to the origin contribute significantly to the cumulants. We then find that the high odd cumulants vanish, while the even ones read
\begin{equation}
 \llangle \hat O_q^{2n} \rrangle \simeq - (2n-1)!\frac{4 \cos(2n \varphi_0)}{|s_0|^{2n}}, \quad n\gg 1,
 \label{eq:4cumu}
\end{equation}
having expressed the zeros in terms of their absolute value, $|s_0|$, and the angle with the real axis, $\varphi_0$. Again, we can determine the zeros from the high cumulants by inverting Eq.~(\ref{eq:4cumu}) for the closest zeros. As described in Appendix~\ref{app:appendixAboutDerivationOfCDW22formula}, we obtain from Eq.~(\ref{eq:4cumu}) the expressions
\begin{widetext}
\begin{equation}
\label{eq:s04}
 |s_0|^4\simeq\frac{(n-1)(n-2) \llangle \hat O_q^{n-2} \rrangle \llangle \hat O_q^{n+2}\rrangle-(n+1)n \llangle \hat O_q^{n} \rrangle^2}{\llangle \hat O_q^{n+4} \rrangle \llangle \hat O_q^{n}\rrangle/(n+3)(n+2)-\llangle \hat O_q^{n+2} \rrangle^2/n(n+1)}
\end{equation}
and
\begin{equation}
  \mathrm{Re}\{s_0^2\} \simeq  \frac{(n-1)(n-2)\llangle \hat O_q^{n-2} \rrangle \llangle \hat O_q^{n+4} \rrangle - (n+3)(n+2) \llangle \hat O_q^{n} \rrangle \llangle \hat O_q^{n+2} \rrangle }{2 \llangle \hat O_q^{n} \rrangle \llangle \hat O_q^{n+4} \rrangle - 2 (n+3)(n+2) \llangle \hat O_q^{n+2} \rrangle^2 / n(n+1)},
  \label{eq:res04}
\end{equation}
which allow us to determine $s_0$ from the high even cumulants, and where we have used that $\mathrm{Re}\{s_0^2\}=|s_0|^2  \cos(2\varphi_0)$.
\end{widetext}

In Fig.~\ref{fig:lyzScaling}(c), we show the zeros for different values of the next-nearest-neighbor interaction and different chain lengths. The nearest-neighbor interaction is set to zero. Moreover, the extrapolation
to the thermodynamic limit is indicated by a line. In Fig.~\ref{fig:lyzScaling}(g), we show the convergence point as a function of the next-nearest-neighbor interaction, and we can again identify the location of the phase transition. For these calculations, we have used cumulant orders of about 10, for which we find that Eq.~(\ref{eq:4cumu}) holds to a good approximation. However, there are also cases, where cumulant orders of 10 are not high enough to extract the closest zeros, and  higher orders may be needed. This situation is discussed in Appendix~\ref{app:appendixAboutDerivationOfCDW22formula} together with the accuracy of the extracted zeros.

Finally, we consider the case of one-third filling, where we expect a CDW with $q = 1/3$ to form for large interaction strengths. In this case, we use the complex order parameter $\hat O_{q } = \sum_{j} \lambda^j\hat n_j$ with $\lambda=e^{i 2\pi/ 3} $. The zeros of the moment-generating function are then typically located as  shown in Fig.~\ref{fig:heatmapPlotSplane}(d), where we notice a rotational symmetry around the origin of the complex plane. This symmetry can be understood from the translational invariance of  the Hamiltonian,
\begin{equation}
 \hat T  \hat \Ham \hat T^\dag= \hat \Ham, 
\end{equation}
where $\hat T$ translates the system by one site. We then find
\begin{equation}
\begin{split}
 \chi_\beta(s) 
 &=  \text{tr}\left[ e^{-\beta \hat T  \hat \Ham \hat T^\dag} e^{s \hat O_{q}} \right]/Z_\beta\\
 &=  \text{tr}\left[    e^{-\beta  \hat\Ham }  e^{s \hat T^\dag \hat O_{q}  \hat T}\right]/Z_\beta=\chi_\beta(s/\lambda),
\end{split}
\end{equation}
where we have made use of the symmetry 
\begin{equation}
 \hat T^\dag  \hat O_{q}  \hat T= \hat O_{q} /\lambda
\end{equation}
of the order parameter. We then see that, if $s_k$ is a  zero of the moment-generating function, then $\lambda s_k$ is also a zero, which explains the symmetry observed in Fig.~\ref{fig:heatmapPlotSplane}(d).

With this symmetry, we can write the cumulants as
\begin{equation}
 \llangle \hat O_q^n \rrangle \simeq - \frac{(n-1)!}{|s_0|^n} (1 + e^{2\pi n i/3} + e^{-2\pi n i/3}), \quad n\gg1,
\end{equation}
which vanishes if $n$ is not a multiple of 3. If it is, we find
\begin{equation}
 \llangle \hat O_q^{3n} \rrangle \simeq -3 \frac{(3n-1)!}{|s_0|^{3n}},
\end{equation}
and we can obtain the zeros from the expression
\begin{equation}
 |s_0|^{3n}  =  3 \frac{(3n-1)!}{|\llangle \hat O^{3n}_{q} \rrangle|}.
\end{equation}
In Fig.~\ref{fig:lyzScaling}(d), we show the extracted zeros for different values of the interaction strengths and different chain lengths. The extrapolation
to the thermodynamic limit is indicated by a line. In Fig.~\ref{fig:lyzScaling}(h), we show the convergence point as a function of the interaction strengths, allowing us to identify the location of the phase transition. 

\section{Phase diagrams}
\label{eq:PDs}

We can now compile the results of our calculations and assemble the phase diagrams in Fig.~\ref{fig:phaseDiagrams}. Figure~\ref{fig:phaseDiagrams}(a) shows the phase diagram at half filling and without next-nearest-neighbor interactions. With increasing nearest-neighbor interactions, a CDW develops with $q=1/2$, and we are able to clearly identify the phase transition into this ordered phase. As an important check of our results, we recover the analytic prediction of a phase transition occurring at $V_{1} = 2$ and $\delta = 0$~\cite{mishra2011phasediagram}. For the SSH model, corresponding to zero interactions, a transition into a topologically nontrivial phase occurs at  $\delta = 0$~\cite{melo2023topological, yahyavi2018variational}; however, it is not associated with a CDW, and we do not detect it here. Figure~\ref{fig:phaseDiagrams}(b) shows the phase diagram at half filling and zero dimerization~\cite{zhuravlev2000breakdown, duan2011bondorder, mishra2011phasediagram, szyniszewski2018fermionic, gotta2021pairing}. Here, we identify two phase boundaries between a disordered phase and CDW phases with $q=1/2$ and $q=1/4$, respectively. The former develops with increasing nearest-neighbor interactions, also in the absence of next-nearest-neighbor interactions. The latter, by contrast, develops with increasing next-nearest-neighbor interactions and does not require nearest-neighbor interactions. Finally, in Fig.~\ref{fig:phaseDiagrams}(c), we show the phase diagram at one third filling~\cite{gotta2021pairing}. In that case, we find a phase transitions into a CDW with $q=1/3$, which only occurs for finite nearest-neighbor and  next-nearest-neighbor interactions.

\section{Conclusions}
\label{sec:conclusion}

Predicting the phase diagram of interacting fermionic systems is a central problem in quantum condensed matter physics. Here, we have implemented a Lee-Yang formalism for interacting fermionic quantum many-body systems featuring charge density wave. We applied it for a fermionic chain with strong interactions to map out its phase diagram. We have shown that the approach enables us to predict a variety of charge density waves that form at different fillings and interaction strengths. Our formalism combines recent developments in Lee-Yang theories of quantum phase transitions with many-body methods based on tensor networks. Specifically, from the high cumulants of the order parameter, we have extracted the dominant zeros of the moment-generating function, and from their convergence points in the thermodynamic limit, we could predict the occurrence of quantum phase transitions. Here, we evaluated the high cumulants using tensor-network methods. Moreover, since high cumulants are, in principle, measurable, the approach may be relevant for future experiments. Our results also provide a starting point for applications of this Lee-Yang formalism to complex quantum many-body systems such as doped Hubbard models featuring charge density waves and other symmetry-broken states. \revision{It would also be interesting to explore whether our Lee-Yang formalism can be extended to topological phase transitions, 
where a system changes its global properties rather than its local order---for example, by going from a trivial insulator to a topological insulator with a non-zero topological invariant. Unlike the conventional phase transitions that we have considered, topological phase transitions often occur without symmetry breaking, and they are reflected in changing Chern numbers or Berry phases. We leave this question as an open problem for future studies.}

\acknowledgements
We acknowledge the computational resources provided by the Aalto Science-IT project and the support from the Research Council of Finland through Grants No.~331342 and No.~358088 and the Finnish Centre of Excellence in Quantum Technology (Grant No.~312299), the Jane and Aatos Erkko Foundation, the Finnish Quantum Flagship and the Japan Society for the Promotion of Science through an Invitational Fellowship for Research in Japan.

\appendix

\section{Tensor-network calculations}

\label{app:mpsdetails}
We use matrix product states as an ansatz for the ground state of the system, and the optimization is done using the density matrix renormalization group (DMRG). For all calculations, we impose periodic boundary conditions. For $\delta = 0$, we use a maximum bond dimension of $\chi_m = 1000$, while for $V_2 = 0$, a maximum bond dimension of $\chi_m = 100$ is sufficient. We check that the calculations are well converged, but encountered convergence issues in region of large $V_2$  at one-third filling, which prevented us from extending our analysis towards higher values of $V_2$. To ensure that we find all ground states, we first run the DMRG algorithm to find one ground state and then run it again until we do not find another orthogonal state within a small energy window. In this way, we ensure that we find all ground states. 

To evaluate the moment-generating function, we use that the matrix product operator for $\exp(s \hat O_q)$ has just bond dimension 1, and it is  therefore straightforward to evaluate. Its local tensor is simply given by
\begin{equation}
 T_{s_j s_j' a_l a_r} = \exp (\coef_j \sigma^z_{s_j s_j'}/2) 1_{a_l a_r},
\end{equation}
where $\hat O_q = \sum_j \coef_j \hat n_j$, $s_j$ and $s_j'$ are the local physical indices, and $a_l$ and $a_r$ are the left and right bond indices. This allows us to map out the moment generating function in the complex plane and compare the results with the zeros extracted from the cumulants.

\begin{figure}
\centering
\includegraphics[width = 1.0\columnwidth]{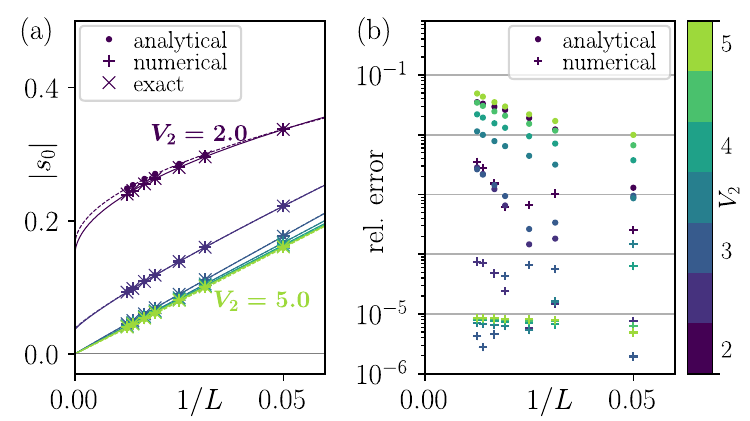}
\caption{\label{fig:accuracyplot} Comparison of analytic and numerical solution. (a)~Extracted zeros based on Eqs.~(\ref{eq:s04}) and (\ref{eq:res04}) and a numerical solution of Eq.~(\ref{eq:4cumu}). The exact results are obtained using a minimum search algorithm as in Fig.~\ref{fig:heatmapPlotSplane}. (b) Relative error.}
\label{fig:comparison}
\end{figure}

\section{Derivation of Eqs.~(\ref{eq:s04},\ref{eq:res04})}
\label{app:appendixAboutDerivationOfCDW22formula}

\begin{figure}
\centering
\includegraphics[width = 0.75\columnwidth]{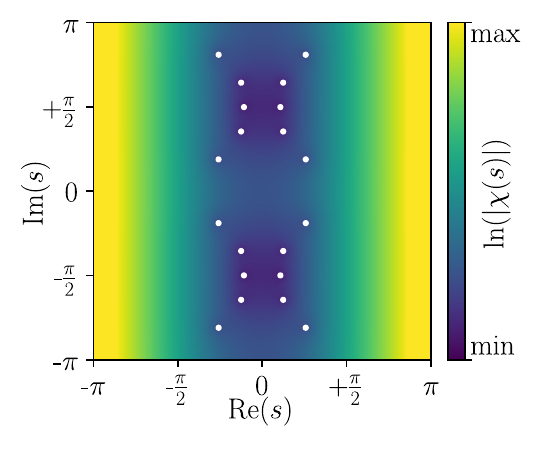}
\caption{\label{fig:problemPlot2} Moment-generating function and complex zeros. Parameters are $q=1/4$, $N=20$, $V_1 = -5$, $V_2 = 2.6$ and $\nu = 1/2$.}
\label{fig:MGFproblem}
\end{figure}

Using Eq.~(\ref{eq:4cumu}), the ratios of two even cumulants read
\begin{equation}
   \kappa_{n-2}^{+}=|s_0|^2\frac{(n-3)!}{(n-1)!}\frac{\cos((n-2)\varphi_0)}{\cos(n\varphi_0)}
\end{equation}
and
\begin{equation}
   \kappa_{n+2}^{-}=\frac{1}{|s_0|^2}\frac{(n+1)!}{(n-1)!}\frac{\cos((n+2)\varphi_0)}{\cos(n\varphi_0)},
\end{equation}
having defined $\kappa_{n}^{\pm}=\llangle\hat O_q^{n} \rrangle/\llangle \hat O_q^{n\pm2}\rrangle$. Next, using that
\begin{equation}
2 \cos(2\varphi) \cos(n\varphi)= \cos((n+2)\varphi) + \cos((n-2) \varphi),
\end{equation}
we can combine these expressions into the equation
\begin{equation}
 \frac{(n+1)!}{(n-3)!}\kappa_{n-2}^{+}= \frac{(n+1)!}{(n-1)!}2\mathrm{Re}\{s_0^2\}-\kappa_{n+2}^{-}|s_0|^4,
\end{equation}
having used that $\mathrm{Re}\{s_0^2\}=|s_0|^2\cos(2\varphi_0)$. By substituting $n$ by $n+2$, we obtain the matrix equation
\begin{equation}
\begin{bmatrix} \frac{(n+1)!}{(n-1)!} & \kappa_{n+2}^{-} \\ \frac{(n+3)!}{(n+1)!} & \kappa_{n+4}^{-}
\end{bmatrix} \begin{bmatrix} 
2\mathrm{Re}\{s_0^2\} \\ 
-|s_0|^4 
\end{bmatrix}=
\begin{bmatrix}
    \frac{(n+1)!}{(n-3)!}\kappa_{n-2}^{+} \\
    \frac{(n+3)!}{(n-1)!} \kappa_{n\phantom{-2}}^{+} 
\end{bmatrix},
\end{equation}
which can then be solved to arrive at Eqs.~(\ref{eq:s04}) and (\ref{eq:res04}). To check these  expressions, we show in Fig.~\ref{fig:comparison} a comparison between the analytic results and a numerical solution of Eq.~(\ref{eq:4cumu}). In Fig.~\ref{fig:comparison}(a), we show results of the two approaches, while Fig.~\ref{fig:comparison}(b) shows the relative error.  Figure~\ref{fig:MGFproblem} shows a situation, where eight zeros almost have the same distance to the origin, and it would require very high cumulant orders to extract the closest ones.

\end{document}